# Liquid–liquid-like phase transitions between high, mid, and low density phases in confined water

Saeed Pourasad[1], Amir Hajibabaei[1], Chang Woo Myung[1], Kwang S. Kim[1*]

[1]Center for Superfunctional Materials, Department of Chemistry and Department of Physics, Ulsan National Institute of Science and Technology (UNIST), 50 UNIST-gil, Ulsan 44919, Republic of Korea.
*e-mail: kimks@unist.ac.kr

**Liquid–liquid phase transition (LLPT) in supercooled water has been a long-standing controversial issue. We show simulation results of real stable first-order phase transitions between high and low density liquid (HDL and LDL)-like structures in confined supercooled water in both positive and negative pressures. These topological phase transitions originate from H-bond network ordering in molecular rotational mode after molecular exchanges are frozen. It is explained by the order parameter-dependent free energy change upon mixing liquid-like and ice-like moieties of H-bond orientations which is governed by their two- to many-body interactions. This unexplored purely H-bond orientation-driven topological phase gives mid-density and stable intermediate mixed-phase with high and low density structures. The phase diagram of supercooled water demonstrate the second and third critical points of water.**

LLPT is the one of the most controversial behaviors of water (*1, 2*), which suggests the existence of high and low density liquid phases that can convert into each other through a 1$^{st}$-order transition at a critical point in the super-cooled region (*3-5*). Various scenarios have been proposed to explain the possible existence of LLPT in water (*3, 6-9*) which discloses positive signs (*10-12*) but yet remains controversial (*13*).

Confined water is widespread in porous materials, biological channels and living cells (*14, 15*). Though the density minimum in the deeply supercooled region was observed in confined water (*16*), there is still no clear report about its LLPT. To search for the LLPT along with HDL and LDL, here we study water in a confined environment. Our system is bilayer water confined between two graphene layers represented with Lennard-Jones (LJ) walls . We performed molecular dynamics simulations of the TIP4P/2005 water-model at various densities ranging from 0.91 to 1.3 gcm$^{-3}$ below 0 ºC (see **Method**s in **Supplementary Materials**). In the deeply supercooled region, the simulation shows that O atoms are practically frozen (quasi-long-ranged). Despite the suggestion that the monomer is globally frozen, highly mobile H atoms provide extra degrees of freedom for the locally unfrozen rotational mode associated with the bond-network. Thus, such a state is not truly ice (because the H-bonding network is frozen in typical ice), but ice-like. Furthermore, the H atoms behave locally liquid-like in the sense that two H atoms in a monomer can be exchanged (which is not allowed in solid state). This unordered rotational mode (with two degrees of freedom in monomer orientation, i.e., parallel and perpendicular to the water monomer plane or to the water layers) can bring about phase transitions. Here, such non-completely frozen states are designated simply as ice-like liquid (L) or preferably ice/liquid-like anisotropic water (W), to facilitate our discussion. Thus, LDL/HDL can be denoted as LDW/HDW, and supercooled liquid as supercooled water (SCW), LLPT as WWPT. In particular, we discuss the results for water molecules with two special densities ($\rho \cong 1.17$ and $0.97$ gcm$^{-3}$) showing WWPT under positive and negative lateral pressures (*P*), respectively.

**Fig. 1** gives the *NVT* simulation results for confined bilayer water systems of $\rho \cong 1.17$ gcm$^{-3}$ under positive lateral pressure (~1.49-1.53 GPa) (**A-C**) and $\rho \cong 0.97$ gcm$^{-3}$ under negative pressure (-*P*=~1.7-2.5 GPa) (**D-F**). The average potential energy (*U*) of the confined water (**Fig. 1A,D**) shows a drop from SCW to an intermediate phase and another drop from the intermediate phase to ice. The bond orientational order parameter $Q_6$ (**Fig. 1B,E**) shows hysteresis loops upon slow cooling/heating processes under both positive and negative pressures. The heat capacity $C_v$ (**Fig. 1C,F**) also show two weak peaks under positive pressures (corresponding to WWPT and ice formation) and similarly two strong peaks under negative pressures. **Fig. 1H** depicts the side view/snapshot of the water configuration under positive pressure, while the difference between SCW and ice structures is not easily visible due to minor changes. However, in **Fig. 1G** the side views/snapshots of water under negative pressure distinguish three different phases (HDW, intermediate HDL-LDL mixture phase, and LDW) in the supercooled region in view of the number of hydrogen bonds (within bond-length ~1.8 Å) and their orientations.

In this regard, we further performed *NPT* simulations to understand more about the intermediate phase. **Fig. 2A-B** show the case for $\rho=1.162\pm0.001$ gcm$^{-3}$ under positive *P* (1.3–1.8 GPa) at 210-194 K and **Fig. 2C-F** present the case for $\rho=0.967\pm0.001$ gcm$^{-3}$ under negative pressure (-*P* =1.1~1.8 GPa) at 210-196K (refer to **Figs. S1-3** for more details). In these two cases, as *T* is lowered from *T*/K$\cong$210, the free energy change *ΔG(P)* shows a phase transition from the SCW to a major portion of HDW (at a higher pressure) and a newly emerging small portion of ice-like LDW (at a lower pressure). The *ΔG(P)* at ~198 ≤ *T*/K ≤ ~205 shows two minima between which one additional shallow minimum [a new mid-density water (MDW) between HDW and LDW (*17*)] appears at $T_0$/K$\cong$200. For *T*/K<200, the major component is ice-like LDW (at the lower pressure). By decreasing the temperature more the liquid-like HDW (at the higher pressure) being of a minor component vanishes and the system turns to the ice phase. The MDW (the red-dotted lines in **Fig. S1C**) is responsible for the unusual hysteresis loop of **Fig. 1B,E** where the <u>transiently stable</u> intermediate states (at *T*/K=200– 205) are indeed located between the ice and SCW states (unlike the typical hysteresis loop that could show direct-jumps from 199 K to 205 K and vice versa with no intermediate state between the two temperatures), which were never observed in previous studies of water.

**Figs. 2B and 2D** show phase diagrams for the supercooled region of the confined water at positive and negative pressures, respectively, along with the characteristic hysteresis loops, based on **Fig. 2A,C** and **Figs. S1-S3**. The two critical temperatures are likely to be associated with the once hypothesized second critical point of water (*3*). **Fig 2E** depicts the side views/snapshots of three intermediate states HDW, MDW, and LDW (which are shown in **Animation Videos 1-3**, respectively). The MDW structure is quite different from HDW and LDW, because it is a transient metastable state (as a barrier state-like structure for the transition between HDW and LDW) which gives much less H-bonds than LDW and HDW. **Fig. 2F** shows the radial distribution function (RDF) of O-H distances ($g_{O-H}$) for solid ice at $T/K=196$, SCW at $T/K=210$, two intermediate states (mainly SCW-like HDW at $T/K=203$ and mainly ice-like LDW at $T/K=198$), and a transient intermediate state of MDW at $T_0/K=200$, under negative pressure.

**Fig. 2G** demonstrates that in the negative pressure ($P=-1.37$ GPa) the free-energy barrier heights at $T/K=200$ with respect to system size ($N$) satisfy the $N^{1/2}$ scaling characteristic of 1$^{st}$-order phase transition (*4*). **Fig. 2H** shows that the $Q_6$ of oxygen atoms in confined water does not vanish even if the system size increases, indicating a partially ordered confined water with the less ordered structure than ice ($Q_6=0.1235$).

Based on this finding, we analyze the system based on Landau phase transition theory (*18*) which we have adapted for water in terms of many body interactions (*19*) between either two components or two interconvertible molecular states or moieties where their fractions are x and 1-x with chemical potentials $\mu_A$ and $\mu_B$ (see **Supplementary Materials: Text, Figs S1-S3** and **Tables S1-S2**). The mixing free energy *ΔG* is the sum of mixing-driven entropic energy change (-*TΔS*), the mixing-driven enthalpy change ($ΔH=ΔH^i+ΔE^\mu$) which is composed of many-body interaction driven energy change ($ΔH^i$), and the chemical potential driven energy change ($ΔE^\mu$) at a given temperature *T*:

*ΔG*= -*TΔS*+*ΔH* = -*TΔS*+*ΔH$^i$*+*ΔE$^\mu$* = *ΔG$^i$*+*ΔE$^\mu$*

where

$ΔG^i = -TΔS+ΔH^i$

$ΔE^\mu = x\mu_A + (1-x)\mu_B.$

$-TΔS = kT\, x \ln x + kT\, (1-x) \ln(1-x)$

where *k* is Boltzmann constant. Here we consider two to six body interactions for the enthalpy change in configurational mixing ($ΔH^i = \Sigma_{ij}\, v_{ij}X_iX_j + \Sigma_{ijk}\, v_{ijk}X_iX_jX_k + \Sigma_{ijkl}\, v_{ijkl}X_iX_jX_kX_l +…$, where $X_i$ is one of two components/states, while $v_{ij}$, $v_{ijk}$, and $v_{ijkl}$ are energies representing 2-, 3- and 4-body interactions in enthalpy) (*20-22*):

$$\Delta H^i = \sum_{m,n} \omega_{mn} x^m (1-x)^n \qquad \text{(for (m+n)-body interactions where m,n} \geq 1\text{)}.$$

The equation ($\Delta G^i = -T\Delta S + \Delta H^i$) provides a symmetric (if $\omega_{mn} = \omega_{nm}$) or asymmetric ($\omega_{mn} \neq \omega_{nm}$) $\Delta G$. Such a $\Delta G$ profile can describe the 1$^{st}$-order phase transition between two interconvertible phase in super-cooled water.

The MDW state at $T/K \cong 200$, upon mixing HDW and LDW moieties, appears as a shallow local minimum in addition to two global minima (HDW, LDW) responsible for the two-step transition mechanism (*17*) in the 6-th order power function of order parameter. This is possible when the effective two-body, four-body and six-body interaction energies between HDW and LDW moieties are attractive, repulsive, and attractive, respectively. This allows the intermediate phase to be in partially mixed HDW-LDW configurations (though within a very narrow temperature region, $T/K=\pm 3$). It can be evidenced from the MDW structure devoid of significant hydrogen bonding, which makes easier the transformation between HDW and HDW, so that the repulsiveness between HDW and LDW (as compared with HDW-HDW or LDW-LDW interactions) becomes attractiveness in 2-body interaction term in the MDW state. If the effective 2-body and 4-body interaction energies between HDW and LDW moieties are repulsive and attractive, respectively, there could appear a spinodal decomposition (*23*) without MDW state (showing the local maximum or barrier of $\Delta G$) which hinders the mixing of two different phase configurations due to their repulsion to each other.

Confined water has the layering effect (*24-25*), which hinders water molecules from leaving the layer and allows them to form a relatively ordered structure. Lowering the temperature of the confined water freezes molecular exchanges first (at ~300/350 K under positive/negative pressure) and then the rotational mode of H atoms against O, i.e., H-bonding orientations. This unordered orientational degree of freedom under frozen molecular exchanges creates a new mixed-phase between SCW and ice (analogous to the hexatic phase phenomenon (*26*) in 2D as an intermediate phase, but with the clear difference that the system here is translationally ordered but orientationally disordered, opposite to the hexatic phase case). This new phase is associated with the anisotropic phenomenon because though the molecular exchanges are frozen, the molecular orientations or bond-networks are unordered in two degrees of freedom of rotation (two orientations parallel and perpendicular to 2D layer or molecular plane). These create two transitions; the first one arises from the transition from SCW to a minimum state comprised of mainly HDW by the parallel orientation ordering, and the second arises from another minimum state comprised of mainly LDW to ice by the perpendicular orientation ordering. Between these two transitions, HDW and LDW are segregated in the spinodal regions below $T_c$ and above $T_l$, but in a small interval of temperature between $T_c$ and $T_l$, the metastable MDW exists between HDW and LDW, allowing characteristic stable hysteresis loops through stable intermediate states without sudden jumps from the initial to final state.

## References

1. Debenedetti, P. G. *J. Phys. Condens. Matter* **15**, R1669–R1726 (2003).
2. Gallo, P. *et al. Chem. Rev.* **116**, 7463–7500 (2016).
3. Poole, P. H. *et al. Nature* **360**, 324–328 (1992).
4. Palmer, J. C. *et al. Nature* **510**, 385–388 (2014).
5. Angell, C. A. *Science* **319**, 582–587 (2008).
6. Poole, P. H. *et al. Phys. Rev. Lett.* **73**, 1632−1635 (1994).
7. Speedy, R. J. *J. Phys. Chem.* **86**, 982–991(1982).
8. Sastry, S., *Phys. Rev. E. Stat. Phys. Plasmas Fluids Relat. Interdiscip. Topics* **53**, 6144–6154 (1996).
9. Tanaka, H. *J. Phys. Condens. Matter* **15**, L703-L711 (2003).
10. Sellberg, J. A. *et al. Nature* **510**, 381–384 (2014).
11. Kim, K. H. *et al. Science* **358**, 1589–1593 (2017).
12. Woutersen, S. *et al. Science* **359**, 1127-1131 (2018).
13. Smart, A. G. *Phys. Today* 22. Aug. 2018.
14. Tros, M. *et al*. *Nat. Commun.* **8**, 1–7 (2017).
15. Cerveny,S. *et al. Chem. Rev.* **116**, 7608-7625 (2016).
16. Liu,D.Z. *Proc. Natl. Acad. Sci. USA* **104**, 9570–9574 (2007).
17. Peng, Y. *et al. Nat. Mater.* **14**, 101–108 (2015).
18. Binder K. *In Phase Transformations in Materials, ed. Kostorz*, G. 239–308 (2001).
19. M. A. Anisimov . *et al. Phys. Rev. X* **8**, 011004 (2018).
20. Kim, K. S.*et al. Chem. Phys. Lett.* **131**, 451−456 (1986).
21. Kim, J.*et al. J. Chem. Phys.* **102**, 839-849 (1995).
22. Kim, J. *et al. Chem. Phys. Lett.* **219**, 243– 246 (1994).
23. Cho, W. J. *et al. Phys. Rev. Lett.* **112**, 157802 (2014).
24. Gao, J.*et al.Phys. Rev. Lett.* **79**, 705–708 (1997).
25. Koga, K. *et al. Nature* **412**, 802–805 (2001).
26. Hajibabaei, A. & Kim, K. S. *Phys. Rev. E.* 99, 022145 (2019).

## Acknowledgements

This work was supported by NRF(2010-0020414) and KISTI(KSC-2020-CRE-0049/KSC-2019-CRE-0253/KSC-2018-CHA-0057/KSC-2017-C3-0081). D.C.Yang and L.Wang are acknowledged for discussion. **Author contributions**: S.P. performed all simulations. K.S.K. performed theoretical analysis. A.H. and C.W.M. helped in simulation and theoretical analyses, respectively. S.P and K.S.K wrote the manuscript. The project was supervised by K.S.K. **Competing interests** : The authors declare that they have no competing interests. **Correspondence and requests** for materials should be addressed to K.S.K. **Data and materials availability:** All data are available in the main text or the supplementary materials.

## SUPPLEMEANTARY MATERIALS

Methods, Text (Many-body interactions driven phase transitions upon mixing), Figs. S1-S3, Tables S1-S2, References (*27-31*).

**Animation** Videos 1-3.

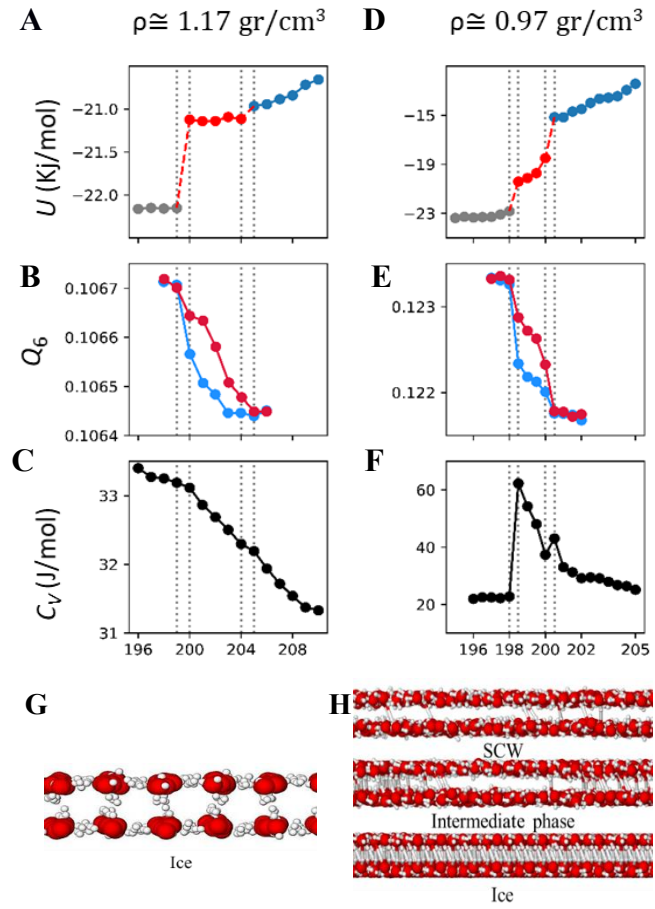

**Fig.1. Temperature dependent features of *NVT* ensemble** for ρ≅1.17 gcm$^{-3}$ under positive pressure (**A-C**) and ρ≅0.97 gcm$^{-3}$ under negative pressure (**D-F**) in the supercooled region of the confined water with LJ walls. **A,D,** Average potential energy (*U*), **B,E,** slow cooling/heating-driven $Q_6$-*T* hysteresis loop, **C,F,** isochoric heat capacity ($C_v$), and **G,H,** Side-views/snapshots of water structures. Under negative pressure, the supercooled water (SCW) phase, intermediate stable phase and ice phase show significant differences in number of hydrogen bonds within bond length ~1.8 Å and bond orientations. In both positive and negative pressures, all the thermodynamic features are similar, while much more conspicuous in the latter.

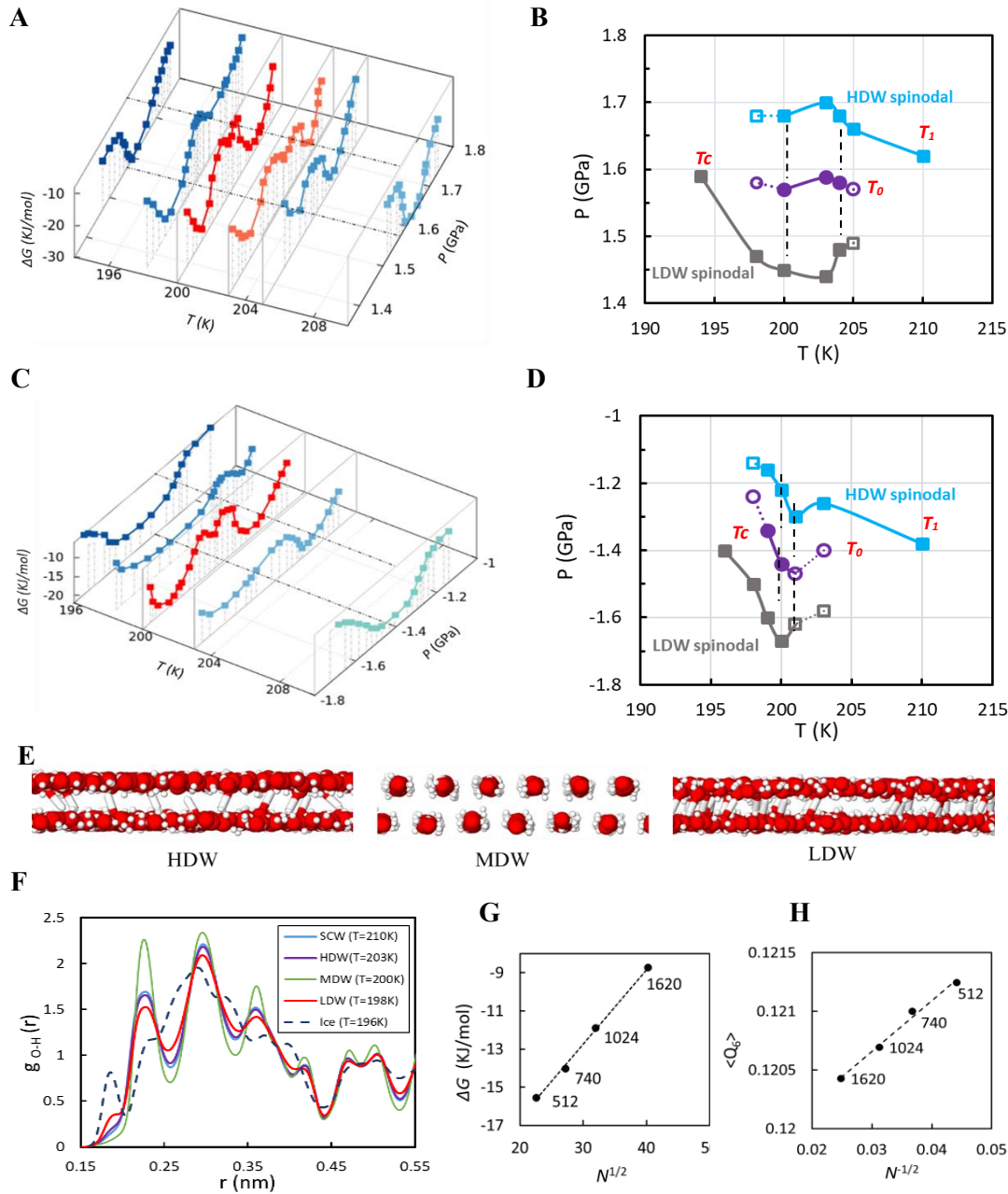

**Fig.2.** *NPT* **simulation features and phase diagram of the confined water under positive (A-B) and negative (C-D) lateral pressures. A,C,** $\Delta G(P)$ profiles at varying temperatures where the asymmetry between left and right minima is reversed around the transient symmetry at $T/K \cong 200$ of the MDW phase between the HDW and LDW phases. **B,D,** Phase diagrams in the supercooled region which show the hysteresis loops, (**Figs. S1C, S2-S3**). **E,F**, Side-views of HDW, MDW and LDW structures and radial distribution function (RDF) of $g_{O-H}$ under negative pressure, which show H-bond features [solid ice at $T/K=196$, ice-like LDW at $T/K \cong 198$, a transient (barrier-state-like) intermediate MDW state at $T_0/K \cong 200$, SCW-like HDW at $T/K \cong 203$, and SCW at $T/K=210$]. **G**, Free energy barrier ($\Delta G^{\ddagger}$) separating the HDW and LDW basins for systems with different number of water molecules ($N$) at negative pressure. **H,** Bond orientation order ($Q_6$) versus $N$ [which confirms the ordered nature for liquid phase of confined bilayer water, though less ordered than the bil-layer ice structure ($Q_6=0.1235$).

7